# Preliminary Results on the Employment Effect of Tourism: A meta-analysis

Georgios Giotis[1]

## Abstract

Unemployment is one of the most important issues in every country. Tourism industry is a dynamic sector which is labor augmented and can create jobs, increase consumption expenditures and offer employment opportunities. In the analysis of this work, the empirical literature on this issue is presented, which indicates that tourism can play a vital and beneficial role to increase employment. This paper uses meta-analysis techniques to investigate the effect of tourism on employment, and finds that the vast majority of the studies reveal a positive relationship. The mean effect of the 36 studies of the meta-sample, using Partial Correlations, is 0.129. This effect is similar to the regression results which is around 0.9, which is positive and statistically significant, clearly indicating that tourism is effective in the creation of jobs and offers employment opportunities. Moreover, evidence of selection bias in the literature in favor of studies which produce positive estimates is found, and once this bias is corrected the true effect is 0.85-0.97.

[1] Corresponding author at: Department of Management Science and Technology, University of Patras, 26334 Patras, Greece, E-mail address: ggiotis@upatras.gr.

1. **Introduction**

The tourism sector in many countries, such as Greece or Portugal, is one of the country's most viable sources of income. This can be partly attributed to the natural morphology of the main land and especially its attractive islands filled with historical antiquity and elegant art while combining alpine and aquatic environment altogether. Not to mention the convenient weather conditions throughout the year which can be exploited for tourism activities able to adapt to each different season. In that sense, being a tourism country, can mean that employment indices could be heavily affected at particular periods of the year. However, this does not always translate to employment growth. In this work, the inherent effect of tourism on the employment is analyzed using the estimates of the empirical studies that have been conducted on this issue across several countries in the globe.

Unemployment rates became one of the most significant matters for the country's economy. Among other sources of income, tourism created some promising opportunities since it is a continuously-growing labor sector. Although sometimes its effects can be temporary, tourism can be a considerable power for a country's employment. When properly organized, regular tourist consumptions are able to provide direct or indirect labor employment opportunities. Either way, small changes in the general economy of a country can be reflected by tourism influence in overall employment for particular time periods.

In this framework, the focus of this work is to investigate whether there is a relationship between tourism and employment, as well as to highlight the potential merit of tourism to employment. More specifically, in this work it is investigated the sign of the employment effect of tourism using a sample of 36 empirical studies which generated 397 estimates. Furthermore, it is attempted to further investigate the degree of this effect. To this end, we perform a meta-analysis on the data collected from these studies so as to extract an in-depth inference of the magnitude this effect takes place.

To our knowledge this is the first approach aiming to investigate the employment effect of tourism using meta-analysis techniques. Hopefully, the results can shed some light to the co-dependence between tourism and employment aiming to alleviate future studies in this respect by exploiting more data.

**2. Literature Review and Meta-Sample**

The literature on the effect of tourism on employment is not relatively large, but it is growing. An extremely thorough and detailed searching has been performed to identify the studies which investigate the relationship between tourism, in general, and employment. More specifically, the searching and coding strategy followed the MAER-NET protocols as outlined in Havránek et al. (2020) by searching REPEC, Google Scholar, Econlit, Scopus, and, additionally, it has been conducted a cited reference search on papers with viable estimates.

The final tourism-employment meta-sample consists of 36 empirical studies which generated 397 estimates. Those studies constitute the meta-sample of the analysis and gave the ability and data to calculate the partial correlations. These studies are presented in table 1 of this section accompanied by some characteristics and the main results on the effect of tourism on employment measures.

It has to be noted that the were many studies which did not allow the generation of the partial correlations, or did not investigate the direct impact of tourism on employment or even important data such as the standard errors or the t-statistics were missing. These studies were several and had to be dropped out of the meta-sample. However, they are related to the impact of tourism on employment and are available by the author upon request with the reason of exclusion of the meta-sample.

Table 1. Included studies of the meta-sample by year of publication

| Study | Author(s) | Year | Publication | Country | Result(s) |
|---|---|---|---|---|---|
| 1 | Kirca and Ozer | 2021 | *Regional Statistics* | Turkey | Tourism demand can contribute to both regional and sectoral employment. |
| 2 | Oguchi and Fen | 2021 | *Technium Social Sciences Journal* | Nigeria | Tourism is positively related to employment rate. |
| 3 | Škrabic Peric *et al.* | 2021 | *Sustainability* | 27 EU member states | UNESCO Heritage Sites have significant positive effects on tourism employment. |
| 4 | Dogru *et al.* | 2020 | *Tourism Management* | 12 major metropolitan statistical areas in the USA | Airbnb supply affects positively employment in all sectors of tourism industries. |
| 5 | Dogru *et al.* | 2020 | *International Journal of Contemporary Hospitality Management* | USA | Hotel investments increase employment. |
| 6 | Min *et al.* | 2020 | *Sustainability* | USA | Tourism affects employment volatility. |
| 7 | González and Surovtseva | 2020 | Barcelona GSE Working Paper Series | Spain | An increase in tourist inflows leads to more employment in the tourism industry for prime-age workers in the short term but does not increase total employment in local economies. |
| 8 | Marques Santos, *et al.* | 2020 | *Publications Office of the European Union* | EU25 | The report displays an estimation for the number of jobs at risk in EU in 2020, as a result of the slowdown of tourism activities. |
| 9 | Storm *et al.* | 2020 | *Regional Studies* | Europe | Hosting Formula 1 races does not produce positive effects. |
| 10 | Ganeshamoorthy | 2019 | *American Journal of Humanities and Social Sciences Research* | Sri Lanka | Tourism would not determine the employment creation in the long-term. |
| 11 | Lanzara and Minerva | 2019 | *Journal of Regional Science* | Italy | Employment in some key service industries is positively related to the inflow of tourists. |
| 12 | Gómez López and Barrón Arreola | 2019 | *Journal of Tourism Analysis* | Mexico | Domestic tourism is the variable with the greatest impact on the generation of direct employment. |
| 13 | Manzoor *et al.* | 2019 | *International Journal of Environmental Research and Public Health* | Pakistan | There is a positive impact of tourism on Pakistan's employment sector. |
| 14 | Wei *et al.* | 2019 | International CHRIE Conference-Refereed Track | China | Tourism employment in China was mainly driven by the development of tourism-related industries. |
| 15 | Ribeiro *et al.* | 2017 | *Tourism Economics* | Brazil | Diversification of the relationship |
| 16 | Vecco and Srakar | 2017 | *European Planning Studies* | Slovenia | Festivals had a positive effect on employment. |
| 17 | Andraz *et al.* | 2016 | *Tourism Economics* | Portugal | There are important regional spillover effects on employment from tourism. |
| 18 | Beneki *et al.* | 2016 | *Tourism Economics* | Greece | The guest capacity of hotels is the strongest factor contributing to employment. |
| 19 | Fang *et al.* | 2016 | *Annals of Tourism Research* | USA | Sharing economy has a positive effect on employment in the tourism industry. |
| 20 | Lim and Zhang | 2016 | *Growth and Change* | Mauritius | Casino expansions exerted a small, positive effect on job growth. |
| 21 | Romao *et al.* | 2016 | *Tourism Economics* | Portugal (Algarve) | The decline of construction activities after 2007 has led to a |

| | | | | | significant increase of regional unemployment. |
|---|---|---|---|---|---|
| 22 | Georgiou | 2015 | SSRN Working Paper | Greece | The growth of tourism sector does not increase the share of the employment of the tertiary sector in the total volume of Greece's employment. |
| 23 | Deloitte and Oxford Economics | 2013 | Deloitte and Oxford Economics for VisitBritain | UK | Spending on job creation has a positive impact for tourism in the UK. |
| 24 | Kadiyali and Kosová | 2013 | *Regional Science and Urban Economics* | USA | The study finds statistically and economically significant employment effects form tourism inflows. |
| 25 | Fortanier and van Wijk | 2010 | *International Business Review* | Mozambique, Tanzania and Ethiopia | The simple scale effects of foreign hotels in east developed countries are positive. |
| 26 | Guisan and Aguayo | 2010 | *Regional and Sectoral Economic Studies* | Spain | Tourism has an important impact on regional employment. |
| 27 | Sun and Wong | 2010 | *Economic Systems Research* | Taiwan | Occupancy rate is positively correlated with jobs-to-sales ratio and employment. |
| 28 | Lazzeretti and Capone | 2009 | *European Planning Studies* | Italy | Local tourist systems affect tourist employment positively. |
| 29 | Cotti | 2008 | *Journal of Gambling Business and Economics* | USA | Counties experience an increase in employment after a casino opens. |
| 30 | Hagn and Maennig | 2008 | *Labour Economics* | Germany | The 1974 Football World Cup held in Germany was not able to generate employment effects in the host cities. |
| 31 | Monchuk | 2007 | *Regional Analysis and Policy* | USA | Counties with a casino opening after 1995 experienced a positive employment effect. |
| 32 | Thompson | 2007 | *Regional Analysis and Policy* | USA | The study identifies a statistically significant relationship between |

| | | | | | basic tourism activity and total employment in Nebraska counties. |
|---|---|---|---|---|---|
| 33 | Aguayo *et al.* | 2006 | 46th Congress of the European Regional Science Association | European Countries | Study shows the positive impact of tourism on the service sector employment. |
| 34 | Papadopoulos and Papanikos | 2005 | *Agricultural Economics* | Greece | The promotion of vinegrowing requires policies that will enlarge the size of vineyards, promote investments in human and physical capital. |
| 35 | Evans and Topoleski | 2002 | NBER Working Paper | USA | In counties where an Indian-owned casino opens, jobs per adult increase by about five percent of the median value. |
| 36 | Guisan and Aguayo | 2002 | *Estudios Económicos Regionales y Sectoriales* | 12 EU countries | Study finds a positive impact of tourism on non-agrarian employment of regions in 12 European Union countries. |

The vast majority of the empirical studies which are included in the meta-sample of the analysis, revealed a positive effect of tourism on employment measures. The theoretical expectations comply with this preliminary result as tourism can create jobs and boost employment at tourism destinations through several ways. However, a further analysis is needed to provide robustness to this positive relationship. In section 3 a graphical analysis of the estimates of the studies that constitute the meta-sample is presented and in section 4 meta-analysis techniques are applied.

## 3. Funnel Graph

The funnel graph is a graphical depiction of the distribution of the estimates of the studies that constitute the meta-sample in relation to the precision i.e., the inverse of the standard error. It is a way to see the sign of the impact and an indirect indication of presence of selection bias. Figure 1 presents the funnel graph of impact of tourism on employment using the PACs (Partial Correlations). The mean value of the effect is 0.129 pointing to a positive impact of tourism on employment of tourism destination.

Figure 1. Funnel Graph on the employment effect of tourism

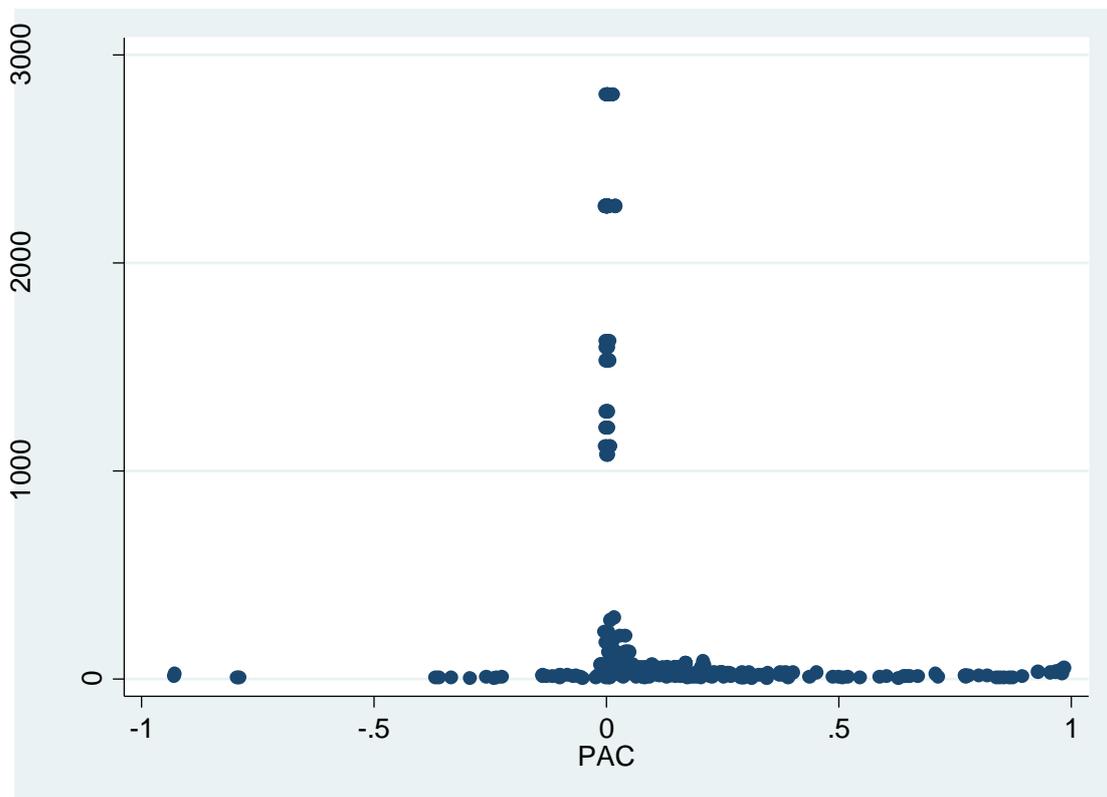

Almost 85% of the partial correlations are positive clearly indicating that tourism plays a beneficial role on the employment of the tourism destination. We observe that most of the PACs gather around the zero-size effect, but we could add that those observations cover the whole area of the positive values and we counted 90 estimates which have precision over 1.000, which is relatively high. However, the percentage of the positive Partial Correlations is 85% which does not leave much to comment on the size of the effect. Only around 15% of the observations of the meta-sample presented a negative impact.

Furthermore, it is important to clarify the existence of publication selection bias and if it corrected, we have to see the genuine, the true effect that remains once this bias is taken into account. Therefore, we continue with the FAT-PET tests to investigate the presence of selection bias and the magnitude of the employment effect of tourism.

4. **FAT-PET and Preliminary Results**

The funnel graph presented in figure 1 provided a clear indication that tourism has a positive effect on employment. However, it is only a graphical depiction of the relationship and further analysis is needed. In this section we apply the OLS (cluster) methodology as well as the Random Effects (REML) and the Method of Moments (MM) to investigate the presence of selection bias and, once this bias is corrected, the true, the genuine effect.

The FAT test is a Funnel Asymmetry Test to discover the presence of bias and the PET test is the Precision Effect Test in order to find the magnitude of the impact.

In the analysis that follows, the PACs (Partial Correlations) are used as size effects. More specifically, the estimates of the employment effect of tourism were collected form each study of the meta-sample and then they were converted into partial correlations, PACs. This is the correlation between tourism and employment. For more details on how to calculate the partial correlations an and their standard errors see Stanley and Doucouliagos (2012).

To test for the presence of publication bias and the precision of the estimated empirical effect we perform the FAT-PET analysis:

$$r_{ij} = \beta_0 + \beta_{se} SE_{ij} + \varepsilon_{ij} \quad (1)$$

where *r* denotes the partial correlation (PAC), *SE* its standard error, and ε is the error term. The FAT tests the hypothesis $H_0: \beta_{se} = 0$. This is a test for censoring of reporting biasness towards the direction of statistically significant results, and the PET tests $H_0: \beta_0 = 0$ to investigate whether there is a genuine effect of tourism on employment once the selection bias is corrected. Tables 2-4 present the regression results using the OLS (cluster), REML (Random Effects Multi-Level) and EB (Empirical Bayes) methods, respectively.

Table 2. OLS (cluster) results

```
. xi: reg  pac  sepac, cluster(study)

Linear regression                               Number of obs =     397
                                                F(  1,    35) =    2.29
                                                Prob > F      =  0.1388
                                                R-squared     =  0.0271
                                                Root MSE      =  .25345

                       (Std. Err. adjusted for 36 clusters in studyid)
```

| pac | Coef. | Robust Std. Err. | t | P>\|t\| | [95% Conf. Interval] | |
|---|---|---|---|---|---|---|
| sepac | .7805757 | .5152644 | 1.51 | 0.139 | -.2654667 | 1.826618 |
| _cons | .0973937 | .0448666 | 2.17 | 0.037 | .0063096 | .1884778 |

Table 3. REML results

```
. metareg pac sepac, wsse(sepac) reml

Meta-regression                                 Number of obs =     397
REML estimate of between-study variance         tau2          = .05019
% residual variation due to heterogeneity       I-squared_res = 98.29%
Proportion of between-study variance explained  Adj R-squared =  6.93%
With Knapp-Hartung modification
```

| pac | Coef. | Std. Err. | t | P>\|t\| | [95% Conf. Interval] | |
|---|---|---|---|---|---|---|
| sepac | 1.178622 | .2690153 | 4.38 | 0.000 | .6497412 | 1.707503 |
| _cons | .0857501 | .0151563 | 5.66 | 0.000 | .055953 | .1155473 |

Table 4. EB results

```
. metareg pac sepac, wsse(sepac) eb

Meta-regression                                      Number of obs  =      397
Empirical Bayes estimate of between-study variance   tau2           =   .05469
% residual variation due to heterogeneity            I-squared_res  =   98.29%
Proportion of between-study variance explained       Adj R-squared  =    4.77%
With Knapp-Hartung modification
------------------------------------------------------------------------------
         pac |      Coef.   Std. Err.      t    P>|t|    [95% Conf. Interval]
-------------+----------------------------------------------------------------
       sepac |   1.154492   .2670161     4.32   0.000    .6295416    1.679442
       _cons |   .0864363    .015205     5.68   0.000    .0565436    .1163291
------------------------------------------------------------------------------
```

The regression results in tables 2-4, indicate that tourism has a positive impact on employment. This result complies with the theoretical expectations as analyzed in the introduction section. The importance now lays in the magnitude of the effect. The mean effect of the PACs of the meta-sample was 0.129, which is similar to the estimations which were generated by table 2 (0.0973), table 3 (0.0857) and table 4 (0.0864). The sign of the effect is undoubtably positive and the magnitude of the impact is around 0.1. However, this result of PACs is not so large, according Doucouliagos (2011), who suggests that partial correlations that are larger than 0.33 can be deemed to be large. Therefore, the effect may be positive, but it is not so large, though.

On the other hand, regarding the dimension of selection bias, in the REML and EB methods, evidence of selection bias in favor of studies which produce positive results is found, while in the OLS (cluster) method, the sign of the *sepac* estimate may be positive but it is not statistically significant. Consequently, we have only indications that there is selection bias in favor of studies which generated positive relationship between tourism and employment, but the empirical evidence is not obvious in all cases.

However, there is a limitation in the analysis. Like any regression model, the estimates of meta-analysis coefficients can become biased when important explanatory variables are omitted. To deal with this issue, the researcher has to incorporate into equation (1) moderator variables that explain variation in tourism-employment relationship. The meta-regression model which should be estimated then (taking account the study heterogeneity) should take the form:

$$r_{ij} = \beta_0 + \beta_{se}\,SE_{ij} + \gamma Z_{ij} + \varepsilon_{ij} \qquad (2)$$

where Z is a vector of variables that reflect modeling differences.

The inclusion of these moderators will provide robustness to the analysis that has already been conducted, and will give light the sources of heterogeneity within the results of the meta-sample. Judging by the coding of the studies, potential sources of heterogeneity could be located in the model characteristics, the country, the dependent and independent variables used in the study and study characteristics related to industry, the sample and methods applied.

5. **Conclusions**

Dealing with unemployment is one of the most important issues in every country. Tourism industry is a dynamic sector which is labor augmented sector and can create jobs, increase consumption expenditures and offer employment opportunities in the tourism destination country. In the analysis presented in this work, we report the empirical literature on this issue which indicates that tourism can play a vital and beneficial role in the increase of employment in a country.

This paper uses meta-analysis techniques to investigate the effect of tourism on employment and finds that the vast majority of the studies revealed a positive relationship. The mean effect of the 36 studies of the meta-sample is 0.129 (Partial Correlations are used). This effect is similar to the regression results, which is around 0.9, clearly indicating that tourism is effective in the creation of jobs and can offer employment opportunities. Moreover, it is found that there is evidence of selection bias in the literature in favor of studies which produced positive estimates and when this bias is corrected the true effect is 0.85-0.97.